\documentclass[prl,aps,twocolumn,amsmath,amssymb,showpacs]{revtex4} %for PRL-style and size feel

\usepackage{graphicx}% Include figure files
\newcommand{\FigureGroesse}{85mm}

\usepackage{hyperref} % hyperrefs

\begin{document}

%%%%%%%%%%%%%%%%%%%%%%%%%%%%%%%%%%%%%%%%%%%%%%%%%%%%%%%%%%%%%%%%%%%%%%%%

\title{
 Observation of Large Atomic-Recoil Induced Asymmetries in Cold Atom
 Spectroscopy
}
\author{C.~W.~Oates, G.~Wilpers, and L.~Hollberg}
\affiliation{Time and Frequency Division,
 National Institute of Standards and Technology,
 325 Broadway,
 Boulder, CO~80305}

\date{\today}

%%%%%%%%%%%%%%%%%%%%%%%%%%%%%%%%%%%%%%%%%%%%%%%%%%%%%%%%%%%%%%%%%%%%%%%%

\begin{abstract}
The atomic recoil effect leads to large (25~\%) asymmetries in simple
spectroscopic investigations of Ca atoms that have been laser-cooled to
$10~\mu$K. Starting with spectra from the more familiar Doppler-broadened
domain, we show how the fundamental asymmetry between absorption and stimulated
emission of light manifests itself when shorter spectroscopic pulses lead to the
Fourier transform regime. These effects occur on frequency scales much larger
than the size of the recoil shift itself, and have not been observed before in
saturation spectroscopy. These results are relevant to state-of-the-art optical
atomic clocks based on freely expanding neutral atoms.
\end{abstract}

%%%%%%%%%%%%%%%%%%%%%%%%%%%%%%%%%%%%%%%%%%%%%%%%%%%%%%%%%%%%%%%%%%%%%%%%

\maketitle
%%%%%%%%%%%%%%%%%%%%%%%%%%%%%%%%%%%%%%%%%%%%%%%%%%%%%%%%%%%%%%%%%%%%%%%%

When we recently started using much colder Ca~atoms for our ultra-high
resolution optical clock spectroscopy, we discovered unexpectedly large
asymmetries in the resulting spectra, which we found to be the direct result of
atomic recoil.\cite{cur03a} Most surprising was that these effects appeared on a
frequency scale more than ten times larger than the size of the recoil effect
itself. In this letter we use a simple saturation spectroscopic configuration
to show how the small effect of atomic recoil can cause these large asymmetries
in the spectra of samples of freely expanding cold atoms. Not only are these
effects of interest from the point of view of basic physics, they also have
important implications for future optical clocks based on laser-cooled neutral
atoms.\cite{cur03a,wil02c}

\begin{figure}
\centerline{\resizebox{\FigureGroesse}{!}{\includegraphics{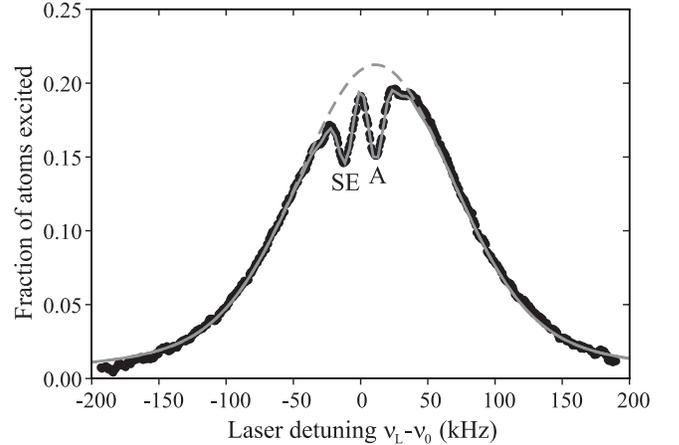}}}
\caption{
Two-pulse saturated absorption spectrum as a function of laser detuning from the
Bohr frequency with a probe pulse duration of $40~\mu$s (circles) along with a
lineshape simulation (solid line) based upon the model described in the text.
Clearly seen are the saturation dips separated by the recoil splitting
superimposed on the Doppler background, with the stimulated emission
component~(`SE') red-shifted and the absorption component~(`A') centered. Shown
for comparison is a simulated ``doubled'' Doppler curve (dashed line). Note that
the much smaller dips on the outside wings of the dips result from the
$(\frac{\sin x}{x})^2$
spectrum of the excitation pulses.
}
\label{DopplerBroadened}
\end{figure}

The effects of atomic recoil have been of considerable interest to laser
spectroscopists for more than 30~years~\cite{kol69} and are the foundation of laser
cooling of atoms. The first experimental evidence that saturation absorption
spectroscopy produced a recoil splitting in the Doppler-free spectra of atomic
and molecular lines was demonstrated by Hall, Bord\'e, and Uehara using a
high-resolution laser spectrometer.\cite{hal76a} Since then there have been several
beautiful demonstrations of recoil splitting via various forms of saturation
spectroscopy (see for example refs.~\cite{bar79,ste92,rie88,bag89}). In all of these studies the
closely spaced recoil components were superimposed on a broad Doppler
background, thereby obscuring the recoil-induced asymmetries inherent in
light-atom interactions. In the work presented here, the width of the Doppler
background is only seven times that of the recoil splitting, thereby clearly
exposing the asymmetric aspect of the recoil components in the Doppler-broadened
regime (see Fig~\ref{DopplerBroadened}).
%%%%%%%%%%%%%%%%%%%%%%%%%%%%%%%%%%%%%%%%%%%%
%%%%%%%%%%%%%%%%%%%%%%%%%%%%%%%%%%%%%%%%%%%%
Moreover, by gradually changing the resolution, we show how this leads
to large spectral asymmetries in the Fourier-transform limit. These effects
differ from asymmetries described in earlier papers, which resulted from
spontaneous emission or collisional quenching.\cite{kol69,hal76a}

As described elsewhere, the recoil doublet can be readily understood from
conservation of momentum and energy.\cite{kol69,hal76a} When an atom initially at rest
absorbs a photon of frequency $\nu$, it recoils with a velocity,
$v_\mathrm{r}=h\nu/mc$,
where $m$ is the atomic mass. For a two-level atom with an energy level spacing
of
$E_\mathrm{0} = h \nu_\mathrm{0}$, this absorption resonance occurs at a frequency,
%%%%%%%%%%%%%%%%%%%%%%%%%%%%%%%%%%%%%%%%%%%%
\begin{equation}
\nu_\mathrm{A} = \nu_\mathrm{0} + \frac{1}{2} \frac{h \nu_\mathrm{0}^2}{m c^2} = \nu_\mathrm{0}
   + \frac{1}{2} \nu_\mathrm{r}.
\label{NuAbsorption}
\end{equation}
%%%%%%%%%%%%%%%%%%%%%%%%%%%%%%%%%%%%%%%%%%%%
Eq.~\ref{NuAbsorption} shows that the incident photon needs to supply energy for
both the atomic excitation and recoil. For an atom at rest but starting in the
excited state, a similar analysis shows that the resonance for stimulated
emission occurs at a frequency
$\nu_\mathrm{SE} = \nu_\mathrm{0} - 1/2\nu_\mathrm{r}$.
This resonance is red-shifted relative
to that for absorption by $\nu_\mathrm{r}$, the splitting of the recoil doublet, which is
typically tens of kilohertz for optical transitions. This frequency splitting
ensures that the alternating absorption and stimulated emission cycles in Rabi
flopping are simultaneously resonant even when the Doppler shift associated with
the atomic recoil is included. Thus, one cannot observe the recoil splitting
with a singly-passed laser beam. Instead, one must use two counter-propagating
laser beams, the usual configuration for saturation spectroscopy. Then the
atomic recoil due to photon absorption from one beam pushes an atom toward the
counter-propagating beam, effectively reversing the sign of the recoil shift.
In this way it is possible to see two distinct sub-Doppler features split by $\nu_\mathrm{r}$,
although this small splitting can only be resolved in ultra-high-resolution experiments.

In this investigation, we take advantage of the capabilities of our optical
clock apparatus \cite{cur03a} to perform the simplest form of saturation spectroscopy. We
excite a laser-cooled ($T = 10~\mu$K) sample of neutral Ca atoms using the closed
657~nm transition between the $^1$S$_0$ ground state and the meta-stable $^3$P$_1$ excited
state, which has a lifetime of $340~\mu$s. With transit-time broadening and
spontaneous emission negligible, we can change the spectroscopic resolution just
by changing the duration of the square probe pulses. Finally, we can excite the
atoms sequentially with laser pulses of equal intensity and frequency detuning.
Sequential excitation greatly simplifies the analysis, since it removes the
possibility of events containing more than two photons from different directions
that can distort the lineshapes \cite{hal76a,bar79,ish94}.

We realize these experimental conditions with the following measurement
cycle.\cite{cur03a} We first load Ca into a magneto-optic trap using the strongly-allowed
423~nm cooling transition. We then turn off the 423~nm light and use a
3-dimensional quenched narrow-line cooling scheme based on the clock transition to
reduce the temperature of the atomic sample ($\sim10^6$~atoms) to less than
10~$\mu$K.\cite{cur03a,bin01a} When the atoms are cold, we switch off the trap and turn on
a greater than 2~G magnetic bias field to perform the spectroscopy on the narrow
$m\!=\!0\!\rightarrow\!m\!=\!0$
clock transition. We then probe the 657~nm clock transition
with pulses derived from a cw~diode laser, which is locked tightly to a narrow
Fabry-Perot cavity fringe (producing a laser linewidth less than 70~Hz). Some
light from this stabilized master laser is used to injection-lock a slave laser
whose output is sent through two acousto-optic modulators to generate the pulses
for the opposing directions. The deflected light is steered into optical fibers
to spatially filter the beams. The beams coupled out of the fibers expand and
are collimated to about 6~mm, so that the atoms see flat wavefronts (radius of
curvature greater than 50~m). As much as 13~mW can be coupled into these beams,
although we adjust the power to yield unit excitation on resonance (commonly
called a $\pi$-pulse in Rabi flopping parlance). For these measurements we
illuminate the atoms with a pulse from one direction, wait 6~$\mu$s (to make
sure the first beam is completely turned off), and then illuminate the atoms
with a pulse from the opposite direction. Finally, the fraction of atoms in the
excited state is measured using a normalized shelving fluorescence detection
technique.\cite{cur03a} We scan the probe frequency ($\nu_\mathrm{L}$) slowly (4~s sweep
time) while continuously repeating the measurement cycle (duration 35~ms) to
generate our spectra as a function of the laser detuning
($\nu_\mathrm{L}-\nu_\mathrm{0}$).

To make a connection with previous experiments, we first consider the
Doppler-broadened regime, for which we choose a pulse duration ($40~\mu$s) such
that the spectroscopic resolution is about 21~kHz. This is less than the
150~kHz (FWHM) Doppler width of the 10~$\mu$K atoms and slightly less than the
23.1~kHz recoil splitting of the 657~nm clock transition. In
Fig.~\ref{DopplerBroadened} we show the resulting excitation spectrum, whose
envelope is primarily determined by the Doppler background. In fact this
envelope is simply twice the height of the curve seen with a single probe pulse,
because for most laser frequencies the two counter-propagating laser beams
excite non-overlapping velocity classes. However, at the absorption resonance
frequency,
$\nu_\mathrm{L}=\nu_\mathrm{A}$
(`A' in Fig.~\ref{DopplerBroadened}), there is not this doubling of the Doppler
background, since the first laser pulse has already excited the majority of
these atoms to the long-lived excited state. Note that the second pulse cannot
de-excite many of these atoms either, since due to atomic recoil they have been
shifted out of resonance. The second resonance (`SE' in
Fig.~\ref{DopplerBroadened}), at
$\nu_\mathrm{L}=\nu_\mathrm{SE}$,
results from atoms that started with a velocity
$v=-v_\mathrm{r}$,
and were thus excited by the first pulse (in the atoms' rest frame,
$\nu_\mathrm{L}=\nu_\mathrm{A}$);
now at rest in the lab frame due to atomic recoil, they are resonant with the
second pulse for stimulated emission, thereby reducing the net fraction of atoms
excited.

The two dips resulting from these resonances are analogous to those seen in
earlier experiments but with one important distinction: due to the small width
of the Doppler background, we can readily see that the dips are asymmetrically
located about the center of the Doppler background. Since the Doppler
background results from absorption, it is naturally centered around the
resonance at
$\nu_\mathrm{L}=\nu_\mathrm{A}$,
coincident with the absorption recoil dip. The dip associated with stimulated
emission is located one recoil frequency
($\nu_\mathrm{r}$)
below the absorption resonance, on the red side of the Doppler curve. It is
important to emphasize that this asymmetry is not a result of the order of the
laser pulses. If we reverse the temporal order, so that the pulse directions are
reversed, we observe the identical lineshape, not its mirror image. Rather,
this asymmetry is a fundamental feature of saturation spectroscopy, though one
that is easily overlooked in experiments with broad Doppler backgrounds. As has
been noted by other observers,\cite{ish94,din94} this asymmetry can lead to undesired
offsets in realizing optical frequency standards based on saturation absorption
since the unperturbed line center of the transition (midway between the recoil
components) is not centered on the background.

\begin{figure}
\centerline{\resizebox{\FigureGroesse}{!}{\includegraphics{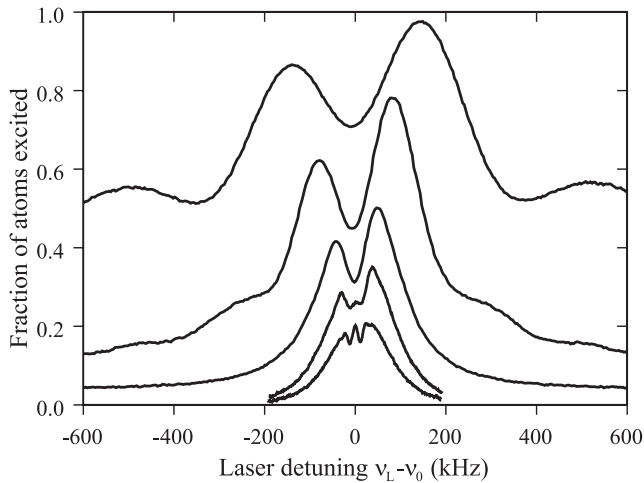}}}
\caption{
Two-pulse spectra taken for probe pulse durations of 40, 20, 10, 6, and 2.6~$\mu$s
(bottom to top). Curves are vertically offset for clarity.
}
\label{PulseShortening}
\end{figure}

We now consider what happens to this spectrum as we broaden the Fourier spectrum
of the probe pulse, not just beyond the recoil splitting but well beyond the
width of the Doppler distribution itself. This regime has not previously been
investigated experimentally, but is becoming important in state-of-the-art
optical atomic clocks based on neutral atoms.\cite{cur03a,wil02c} We access this regime by
reducing the duration of the probe pulses (while maintaining the pulse area to
keep the excitation probability constant). In Fig.~\ref{PulseShortening}, we show a set of
spectroscopic lineshapes taken over probe durations ranging from 40~$\mu$s down to
2.6~$\mu$s.
%%%%%%%%%%%%%%%%%%%%%%%%%%%%%%%%%%%%%%%%%%%%
%%%%%%%%%%%%%%%%%%%%%%%%%%%%%%%%%%%%%%%%%%%%
Note that the fraction of atoms excited grows with decreasing resolution as our
probe spectrum covers a larger fraction of the velocity distribution. For the
shortest pulse length, the Fourier transform of the probe pulse has a spectral
width approximately 2.5~times that of the Doppler distribution. As we see,
changing the resolution from 21~kHz to 42~kHz (20~$\mu$s pulse duration) begins to
obscure the recoil splitting, but the large asymmetry persists. As we move to
85~kHz resolution (10~$\mu$s pulse duration), the recoil splitting is no longer
visible as the two dips merge into a single dip centered at $\nu_\mathrm{0}$,
but the spectrum now appears to consist of two peaks whose separation is
determined by the spectroscopic resolution and whose amplitudes differ by more
than 25~\%\,! The asymmetry persists for even the lowest resolution, where the
maxima are separated by 280~kHz, more than 10~times that of the recoil splitting
itself.

Examination of Fig.~\ref{PulseShortening} provides an intuitive picture of how
such a small effect can cause such large asymmetries. In the Fourier-transform
regime we find the unusual spectroscopic condition where the widths of both the
envelope and the dip are determined almost solely by the probe time, but are
slightly offset from one another. This small offset leads to the envelope
asymmetry by causing the dip to intersect the Doppler envelope at different
heights on the two sides. Thus, the size of the asymmetry is related to the
slope of the envelope multiplied by the size of the recoil effect.

Alternatively, we can think of this spectrum as resulting from two
contributions: one symmetric (due solely to absorption) and one asymmetric (due
to stimulated emission by the second pulse). This framework allows a simple
model to describe our spectra well (an exact theoretical treatment of coherent
saturation spectroscopy including multi-photon effects has been developed by
Bord\'e and co-workers\cite{ish94,bor84}. We start by considering the excitation spectrum
resulting from the first (idealized) square $\pi$-pulse of duration $T$ and Rabi
frequency $\Omega$ illuminating a sample of ground-state atoms initially at rest. This
yields the well-known Rabi spectrum~\cite{rab39} for the excitation probability:
%%%%%%%%%%%%%%%%%%%%%%%%%%%%%%%%%%%%%%%%%%%%
\begin{equation}
P(\Delta) = \frac{\Omega^2T^2}{4}
\left(
\frac{\sin\!\left[\sqrt{\Omega^2+\Delta^2}\ T/2\right]}{\left[\sqrt{\Omega^2+\Delta^2}\ T/2\right]}
\right)^2,
\label{ExcitationProbability}
\end{equation}
%%%%%%%%%%%%%%%%%%%%%%%%%%%%%%%%%%%%%%%%%%%%
where
$\Delta = \nu_\mathrm{L} - \nu_\mathrm{A}$
is the laser detuning from the absorption resonance. Illuminating the atoms with
a square {$\pi$-pulse} from the opposite direction gives two sets of atoms to
consider. First, there are atoms that remained in the ground state after the
first pulse (a fraction equal to
$1- P(\Delta)$);
they will be resonant with a second pulse at frequency
$\nu_\mathrm{L}=\nu_\mathrm{A}$,
so the above equation applies to these atoms as well. Second, the atoms that
were excited by the first pulse (a fraction equal to $P(\Delta)$) now have a
velocity $v_\mathrm{r}$ toward the second counter-propagating laser beam. The
associated Doppler effect will shift the stimulated emission resonance down by
one recoil, so these atoms will be resonant with light at frequency
$\nu_\mathrm{SE}'
= \nu_\mathrm{SE} - \nu_\mathrm{r}
= \nu_\mathrm{0} - \frac{3}{2}\nu_\mathrm{r}$,
or
$\nu_\mathrm{SE}'
= \nu_\mathrm{A}-2\nu_\mathrm{r}$.
The fraction of these atoms transferred back to the ground state can then be
derived from the same probability function $P$, but with the argument
$\Delta'
= \nu_\mathrm{L} - \nu_\mathrm{SE}'
= \Delta+2\nu_\mathrm{r}$.

\begin{figure}
\centerline{\resizebox{\FigureGroesse}{!}{\includegraphics{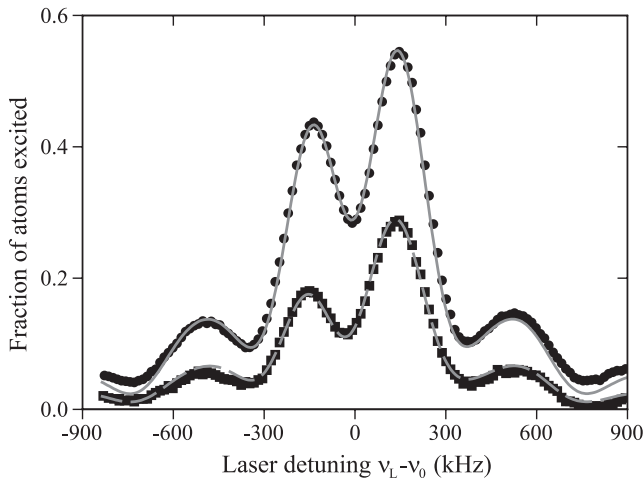}}}
\caption{
Two-pulse spectrum (circles) resulting from probe pulses of 2.6~$\mu$s duration along with the corresponding simulation (solid line). Also shown are the spectrum (squares) and simulation (dashed line) resulting when a blue heating pulse is inserted between red probe pulses to suppress the absorption-based recoil component (see text).
}
\label{FourierTransformed}
\end{figure}

To find the total fraction in the excited state after two pulses, we simply add
the two contributions:
%%%%%%%%%%%%%%%%%%%%%%%%%%%%%%%%%%%%%%%%%%%%
\begin{equation}
P_\mathrm{final} = P(\Delta) \left[ 1-P(\Delta) \right] +
     \left[ 1-P(\Delta+2\nu_\mathrm{r}) \right] P(\Delta).
\label{FinalExcitationProbability}
\end{equation}
%%%%%%%%%%%%%%%%%%%%%%%%%%%%%%%%%%%%%%%%%%%%
The first product in this expression gives the ground-state contribution and is
symmetric (both pulses are resonant at the same frequency). The second term
gives the excited-state contribution, but it is asymmetric due to the offset of
$2\nu_\mathrm{r}$ between arguments of the multiplicands, thus yielding a net asymmetry for
$P_\mathrm{final}$. We can easily connect this model with experiment by including the
initial velocity distribution (via the detuning) and the unequal laser
intensities seen by the atoms due to the spatial distribution of the atoms in
the laser mode (via the Rabi frequency). The free parameters in our simulation
were overall signal amplitude (imperfect normalization required a 10\,-\,15~\%
reduction in signal size) and atomic cloud size (which we fixed at the same
value for all simulations). We measured the velocity distributions and the
laser mode size separately and used the resulting values for the simulations.
We see good agreement over a variety of probe resolutions (simulations are shown
as solid lines in Figs.~\ref{DopplerBroadened} and~\ref{FourierTransformed}),
although we see small differences in the wings at the lowest resolutions,
%%%%%%%%%%%%%%%%%%%%%%%%%%%%%%%%%%%%%%%%%%%%
%%%%%%%%%%%%%%%%%%%%%%%%%%%%%%%%%%%%%%%%%%%%
most likely resulting from the non-ideal square pulses used in the experiment.

Interestingly, we can isolate the second product in the expression for $P_\mathrm{final}$ experimentally by using recoil suppression.\cite{rie88,din94} In this case we suppress the ground-state (absorption) contribution by illuminating the atoms with a resonant 423~nm pulse (duration 20~$\mu$s) after the first red probe pulse but before the second red pulse. This heats the atoms in the ground state to the point where excitation by the second pulse produces a nearly flat Doppler background, upon which sits the contribution from the atoms shelved in the excited state. Fig.~\ref{FourierTransformed} shows that the feature associated with stimulated emission is fully responsible for the lineshape asymmetry, as the absolute peak height differences are virtually identical. We emphasize that this asymmetric envelope persists in four-pulse Bord\'e-Ramsey saturation spectroscopy~\cite{bor84} in the Fourier-transform regime, which is used in ultra-high resolution studies and for cold atom optical clocks.\cite{cur03a,wil02c} However, the good agreement between theory and experiments give us good confidence that these effects will not limit the accuracy of these clocks.

In summary, we have used ultra-cold two-level atoms and ultra-high resolution spectroscopy to probe the atomic recoil structure unique to saturated absorption spectroscopy at an unprecedented level. The resulting spectra clearly reveal the fundamental asymmetry in the location of recoil components on the Doppler background. In addition we have shown how this asymmetry in the recoil frequencies leads to large amplitude asymmetries in the Fourier transform limit, a regime that has not previously been investigated but is important for state-of-the-art precision metrology. Extending these studies to a regime with Ca atoms laser cooled to sub-recoil temperatures, as demonstrated by Curtis et~al.~\cite{cur03a}, would allow us to individually address the recoil components via laser detuning. This could enhance the sensitivity of precision measurements such as that of the photon recoil or other atom interferometry experiments~\cite{wei93,sna98,gup02}.

\begin{acknowledgments}
We thank J. Bergquist and J. Kitching for their insightful comments and J. Hall for illuminating discussions. This work was supported in part by NASA and ONR-MURI. G. W. acknowledges support from the Alexander von Humboldt Foundation.

Work of a US government agency; not subject to copyright.
\end{acknowledgments}
%%%%%%%%%%%%%%%%%%%%%%%%%%%%%%%%%%%%%%%%%%%%%%%%%%%%%%%%%%%%%%%%%%%%%%%%

%\bibliography{//Trumpet/847Div/OFM/Calcium/TexBib/texbi431}

\begin{thebibliography}{16}
\expandafter\ifx\csname natexlab\endcsname\relax\def\natexlab#1{#1}\fi
\expandafter\ifx\csname bibnamefont\endcsname\relax
 \def\bibnamefont#1{#1}\fi
\expandafter\ifx\csname bibfnamefont\endcsname\relax
 \def\bibfnamefont#1{#1}\fi
\expandafter\ifx\csname citenamefont\endcsname\relax
 \def\citenamefont#1{#1}\fi
\expandafter\ifx\csname url\endcsname\relax
 \def\url#1{\texttt{#1}}\fi
\expandafter\ifx\csname urlprefix\endcsname\relax\def\urlprefix{URL }\fi
\providecommand{\bibinfo}[2]{#2}
\providecommand{\eprint}[2][]{\url{#2}}

\bibitem[{\citenamefont{Curtis et~al.}(2003)\citenamefont{Curtis, Oates, and
 Hollberg}}]{cur03a}
\bibinfo{author}{\bibfnamefont{E.~A.} \bibnamefont{Curtis}},
 \bibinfo{author}{\bibfnamefont{C.~W.} \bibnamefont{Oates}}, \bibnamefont{and}
 \bibinfo{author}{\bibfnamefont{L.}~\bibnamefont{Hollberg}},
 \bibinfo{journal}{J. Opt. Soc. Am.~B} \textbf{\bibinfo{volume}{20}},
 \bibinfo{pages}{977} (\bibinfo{year}{2003}).

\bibitem[{\citenamefont{Wilpers et~al.}(2002)\citenamefont{Wilpers, Binnewies,
 Degenhardt, Sterr, Helmcke, and Riehle}}]{wil02c}
\bibinfo{author}{\bibfnamefont{G.}~\bibnamefont{Wilpers}},
 \bibinfo{author}{\bibfnamefont{T.}~\bibnamefont{Binnewies}},
 \bibinfo{author}{\bibfnamefont{C.}~\bibnamefont{Degenhardt}},
 \bibinfo{author}{\bibfnamefont{U.}~\bibnamefont{Sterr}},
 \bibinfo{author}{\bibfnamefont{J.}~\bibnamefont{Helmcke}}, \bibnamefont{and}
 \bibinfo{author}{\bibfnamefont{F.}~\bibnamefont{Riehle}},
 \bibinfo{journal}{Phys. Rev. Lett.} \textbf{\bibinfo{volume}{89}},
 \bibinfo{pages}{230801} (\bibinfo{year}{2002}).

\bibitem[{\citenamefont{Kol{\textquoteright}chenko
 et~al.}(1969)\citenamefont{Kol{\textquoteright}chenko, Rautian, and
 Sokolovski{\u{\i}}}}]{kol69}
\bibinfo{author}{\bibfnamefont{A.~P.}
 \bibnamefont{Kol{\textquoteright}chenko}},
 \bibinfo{author}{\bibfnamefont{S.~G.} \bibnamefont{Rautian}},
 \bibnamefont{and} \bibinfo{author}{\bibfnamefont{R.~I.}
 \bibnamefont{Sokolovski{\u{\i}}}}, \bibinfo{journal}{Soviet Physics JETP}
 \textbf{\bibinfo{volume}{28}}, \bibinfo{pages}{986} (\bibinfo{year}{1969}).

\bibitem[{\citenamefont{Hall et~al.}(1976)\citenamefont{Hall, Bord\'e, and
 Uehara}}]{hal76a}
\bibinfo{author}{\bibfnamefont{J.~L.} \bibnamefont{Hall}},
 \bibinfo{author}{\bibfnamefont{C.~J.} \bibnamefont{Bord\'e}},
 \bibnamefont{and} \bibinfo{author}{\bibfnamefont{K.}~\bibnamefont{Uehara}},
 \bibinfo{journal}{Phys. Rev. Lett.} \textbf{\bibinfo{volume}{37}},
 \bibinfo{pages}{1339} (\bibinfo{year}{1976}).

\bibitem[{\citenamefont{Barger et~al.}(1979)\citenamefont{Barger, Bergquist,
 English, and Glaze}}]{bar79}
\bibinfo{author}{\bibfnamefont{R.~L.} \bibnamefont{Barger}},
 \bibinfo{author}{\bibfnamefont{J.~C.} \bibnamefont{Bergquist}},
 \bibinfo{author}{\bibfnamefont{T.~C.} \bibnamefont{English}},
 \bibnamefont{and} \bibinfo{author}{\bibfnamefont{D.~J.} \bibnamefont{Glaze}},
 \bibinfo{journal}{Appl. Phys. Lett.} \textbf{\bibinfo{volume}{34}},
 \bibinfo{pages}{850} (\bibinfo{year}{1979}).

\bibitem[{\citenamefont{Sterr et~al.}(1992)\citenamefont{Sterr, Sengstock,
 M{\"u}ller, Bettermann, and Ertmer}}]{ste92}
\bibinfo{author}{\bibfnamefont{U.}~\bibnamefont{Sterr}},
 \bibinfo{author}{\bibfnamefont{K.}~\bibnamefont{Sengstock}},
 \bibinfo{author}{\bibfnamefont{J.~H.} \bibnamefont{M{\"u}ller}},
 \bibinfo{author}{\bibfnamefont{D.}~\bibnamefont{Bettermann}},
 \bibnamefont{and} \bibinfo{author}{\bibfnamefont{W.}~\bibnamefont{Ertmer}},
 \bibinfo{journal}{Appl. Phys.~B} \textbf{\bibinfo{volume}{54}},
 \bibinfo{pages}{341} (\bibinfo{year}{1992}).

\bibitem[{\citenamefont{Riehle et~al.}(1988)\citenamefont{Riehle, Ishikawa, and
 Helmcke}}]{rie88}
\bibinfo{author}{\bibfnamefont{F.}~\bibnamefont{Riehle}},
 \bibinfo{author}{\bibfnamefont{J.}~\bibnamefont{Ishikawa}}, \bibnamefont{and}
 \bibinfo{author}{\bibfnamefont{J.}~\bibnamefont{Helmcke}},
 \bibinfo{journal}{Phys. Rev. Lett.} \textbf{\bibinfo{volume}{61}},
 \bibinfo{pages}{2092} (\bibinfo{year}{1988}).

\bibitem[{\citenamefont{Bagayev et~al.}(1989)\citenamefont{Bagayev, Baklanov,
 Chebotayev, and Dychkov}}]{bag89}
\bibinfo{author}{\bibfnamefont{S.~N.} \bibnamefont{Bagayev}},
 \bibinfo{author}{\bibfnamefont{A.~E.} \bibnamefont{Baklanov}},
 \bibinfo{author}{\bibfnamefont{V.~P.} \bibnamefont{Chebotayev}},
 \bibnamefont{and} \bibinfo{author}{\bibfnamefont{A.~S.}
 \bibnamefont{Dychkov}}, \bibinfo{journal}{Appl. Phys.~B}
 \textbf{\bibinfo{volume}{48}}, \bibinfo{pages}{31} (\bibinfo{year}{1989}).

\bibitem[{\citenamefont{Ishikawa et~al.}(1994)\citenamefont{Ishikawa, Riehle,
 Helmcke, and Bord{\'e}}}]{ish94}
\bibinfo{author}{\bibfnamefont{J.}~\bibnamefont{Ishikawa}},
 \bibinfo{author}{\bibfnamefont{F.}~\bibnamefont{Riehle}},
 \bibinfo{author}{\bibfnamefont{J.}~\bibnamefont{Helmcke}}, \bibnamefont{and}
 \bibinfo{author}{\bibfnamefont{C.~J.} \bibnamefont{Bord{\'e}}},
 \bibinfo{journal}{Phys. Rev.~A} \textbf{\bibinfo{volume}{49}},
 \bibinfo{pages}{4794} (\bibinfo{year}{1994}).

\bibitem[{\citenamefont{Binnewies et~al.}(2001)\citenamefont{Binnewies,
 Wilpers, Sterr, Riehle, Helmcke, Mehlst{\ a}ubler, Rasel, and
 Ertmer}}]{bin01a}
\bibinfo{author}{\bibfnamefont{T.}~\bibnamefont{Binnewies}},
 \bibinfo{author}{\bibfnamefont{G.}~\bibnamefont{Wilpers}},
 \bibinfo{author}{\bibfnamefont{U.}~\bibnamefont{Sterr}},
 \bibinfo{author}{\bibfnamefont{F.}~\bibnamefont{Riehle}},
 \bibinfo{author}{\bibfnamefont{J.}~\bibnamefont{Helmcke}},
 \bibinfo{author}{\bibfnamefont{T.~E.} \bibnamefont{Mehlst{\"a}ubler}},
 \bibinfo{author}{\bibfnamefont{E.~M.} \bibnamefont{Rasel}}, \bibnamefont{and}
 \bibinfo{author}{\bibfnamefont{W.}~\bibnamefont{Ertmer}},
 \bibinfo{journal}{Phys. Rev. Lett.} \textbf{\bibinfo{volume}{87}},
 \bibinfo{pages}{123002} (\bibinfo{year}{2001}).

\bibitem[{\citenamefont{Dingler et~al.}(1994)\citenamefont{Dingler, Rieger,
 Sengstock, Sterr, and Ertmer}}]{din94}
\bibinfo{author}{\bibfnamefont{F.~E.} \bibnamefont{Dingler}},
 \bibinfo{author}{\bibfnamefont{V.}~\bibnamefont{Rieger}},
 \bibinfo{author}{\bibfnamefont{K.}~\bibnamefont{Sengstock}},
 \bibinfo{author}{\bibfnamefont{U.}~\bibnamefont{Sterr}}, \bibnamefont{and}
 \bibinfo{author}{\bibfnamefont{W.}~\bibnamefont{Ertmer}},
 \bibinfo{journal}{Opt. Commun.} \textbf{\bibinfo{volume}{110}},
 \bibinfo{pages}{99} (\bibinfo{year}{1994}).

\bibitem[{\citenamefont{Bord{\'e} et~al.}(1984)\citenamefont{Bord{\'e},
 Salomon, Avrillier, {Van~Lerberghe}, Br{\'e}ant, Bassi, and Scoles}}]{bor84}
\bibinfo{author}{\bibfnamefont{C.~J.} \bibnamefont{Bord{\'e}}},
 \bibinfo{author}{\bibfnamefont{C.}~\bibnamefont{Salomon}},
 \bibinfo{author}{\bibfnamefont{S.}~\bibnamefont{Avrillier}},
 \bibinfo{author}{\bibfnamefont{A.}~\bibnamefont{{Van~Lerberghe}}},
 \bibinfo{author}{\bibfnamefont{C.}~\bibnamefont{Br{\'e}ant}},
 \bibinfo{author}{\bibfnamefont{D.}~\bibnamefont{Bassi}}, \bibnamefont{and}
 \bibinfo{author}{\bibfnamefont{G.}~\bibnamefont{Scoles}},
 \bibinfo{journal}{Phys. Rev.~A} \textbf{\bibinfo{volume}{30}},
 \bibinfo{pages}{1836} (\bibinfo{year}{1984}).

\bibitem[{\citenamefont{Rabi et~al.}(1939)\citenamefont{Rabi, Millman, Kusch,
 and Zacharias}}]{rab39}
\bibinfo{author}{\bibfnamefont{I.~I.} \bibnamefont{Rabi}},
 \bibinfo{author}{\bibfnamefont{S.}~\bibnamefont{Millman}},
 \bibinfo{author}{\bibfnamefont{P.}~\bibnamefont{Kusch}}, \bibnamefont{and}
 \bibinfo{author}{\bibfnamefont{J.~R.} \bibnamefont{Zacharias}},
 \bibinfo{journal}{Phys. Rev.} \textbf{\bibinfo{volume}{55}},
 \bibinfo{pages}{526} (\bibinfo{year}{1939}).

\bibitem[{\citenamefont{Weiss et~al.}(1993)\citenamefont{Weiss, Young, and
 Chu}}]{wei93}
\bibinfo{author}{\bibfnamefont{D.~S.} \bibnamefont{Weiss}},
 \bibinfo{author}{\bibfnamefont{B.~C.} \bibnamefont{Young}}, \bibnamefont{and}
 \bibinfo{author}{\bibfnamefont{S.}~\bibnamefont{Chu}},
 \bibinfo{journal}{Phys. Rev. Lett.} \textbf{\bibinfo{volume}{70}},
 \bibinfo{pages}{2706} (\bibinfo{year}{1993}).

\bibitem[{\citenamefont{Snadden et~al.}(1998)\citenamefont{Snadden, McGuirk,
 Bouyer, Haritos, and Kasevich}}]{sna98}
\bibinfo{author}{\bibfnamefont{M.~J.} \bibnamefont{Snadden}},
 \bibinfo{author}{\bibfnamefont{J.~M.} \bibnamefont{McGuirk}},
 \bibinfo{author}{\bibfnamefont{P.}~\bibnamefont{Bouyer}},
 \bibinfo{author}{\bibfnamefont{K.~G.} \bibnamefont{Haritos}},
 \bibnamefont{and} \bibinfo{author}{\bibfnamefont{M.~A.}
 \bibnamefont{Kasevich}}, \bibinfo{journal}{Phys. Rev. Lett.}
 \textbf{\bibinfo{volume}{81}}, \bibinfo{pages}{971} (\bibinfo{year}{1998}).

\bibitem[{\citenamefont{Gupta et~al.}(2002)\citenamefont{Gupta, Dieckmann,
 Hadzibabic, and Pritchard}}]{gup02}
\bibinfo{author}{\bibfnamefont{S.}~\bibnamefont{Gupta}},
 \bibinfo{author}{\bibfnamefont{K.}~\bibnamefont{Dieckmann}},
 \bibinfo{author}{\bibfnamefont{Z.}~\bibnamefont{Hadzibabic}},
 \bibnamefont{and} \bibinfo{author}{\bibfnamefont{D.~E.}
 \bibnamefont{Pritchard}}, \bibinfo{journal}{Phys. Rev. Lett.}
 \textbf{\bibinfo{volume}{89}}, \bibinfo{pages}{140401}
 (\bibinfo{year}{2002}).

\end{thebibliography}
%\bibliographystyle{//Trumpet/847Div/OFM/Calcium/TexBib/apsrev}

%%%%%%%%%%%%%%%%%%%%%%%%%%%%%%%%%%%%%%%%%%%%%%%%%%%%%%%%%%%%%%%%%%%%%%%%
%might wanna put the bibliography entries here

%%%%%%%%%%%%%%%%%%%%%%%%%%%%%%%%%%%%%%%%%%%%%%%%%%%%%%%%%%%%%%%%%%%%%%%%

%might wanna put figures here

%%%%%%%%%%%%%%%%%%%%%%%%%%%%%%%%%%%%%%%%%%%%%%%%%%%%%%%%%%%%%%%%%%%%%%%%

\end{document}